\documentclass[journal]{IEEEtran}

\usepackage{cite}
\usepackage{amsmath,amssymb}
\usepackage{booktabs}
\usepackage{algorithmic}
\usepackage{graphicx}
\usepackage{textcomp}
\usepackage{xcolor}
\usepackage{cite,caption}
\usepackage{fleqn}
\usepackage{eufrak}
\usepackage{amsfonts}
\usepackage{amsthm}
\usepackage{algorithm,algorithmic}
\allowdisplaybreaks
\usepackage{graphicx}
\usepackage{textcomp}
\usepackage{bm}
\usepackage{pifont}
\usepackage{bbm}
\usepackage{stfloats}
\usepackage{indentfirst} 
\usepackage{hyperref}
\usepackage{color}
\usepackage{url}
\usepackage{array}
\usepackage{latexsym}
\usepackage{mathrsfs}
\def\BibTeX{{\rm B\kern-.05em{\sc i\kern-.025em b}\kern-.08em
    T\kern-.1667em\lower.7ex\hbox{E}\kern-.125emX}}
\begin{document}

\title{ Handoff-Aware Distributed Computing in High Altitude Platform Station (HAPS)--Assisted Vehicular Networks}

\author{Qiqi~Ren,
	Omid~Abbasi,~\IEEEmembership{Senior~Member,~IEEE,}
	Gunes~Karabulut~Kurt,~\IEEEmembership{Senior~Member,~IEEE,}
	Halim~Yanikomeroglu,~\IEEEmembership{Fellow,~IEEE,}
	and~Jian~Chen,~\IEEEmembership{Member,~IEEE}
	\thanks{This work is supported in part by the National Natural Foundation of
		China (Grant No. 61771366), and in part by Huawei Canada (Corresponding Author: Jian Chen).}
	\thanks{Q. Ren is with the State Key Laboratory of Integrated Service Networks, Xidian University, Xi' an 710071, Shaanxi, P. R. China and the Non-Terrestrial Networks(NTN) Lab, Department of Systems and Computer Engineering, Carleton University, Ottawa, ON, Canada (email: renqiqi5277@gmail.com) .}
	\thanks{J. Chen is with the State Key Laboratory of Integrated Service Networks, Xidian University, Xi' an 710071, Shaanxi, P. R. China (email: jianchen@mail.xidian.edu.cn).}
	\thanks{O. Abbasi and H. Yanikomeroglu are with the Non-Terrestrial Networks (NTN) Lab,  
		Department of of Systems and Computer Engineering, Carleton University, Ottawa, ON, Canada (email: \{omidabbasi, halim\}@sce.carleton.ca).}
	\thanks{G. Karabulut Kurt is with the Department of Electrical Engineering,  Polytechnique Montr\'eal, Montr\'eal, Canada (e-mail: gunes.kurt@polymtl.ca)}}

\maketitle
\begin{abstract}
   Distributed computing enables Internet of vehicle (IoV) services by collaboratively utilizing the computing resources from the network edge and the vehicles. However, the computing interruption issue caused by frequent edge network handoffs, and a severe shortage of computing resources are two problems in providing IoV services. High altitude platform station (HAPS) computing can be a promising addition to existing distributed computing frameworks because of its wide coverage and strong computational capabilities. In this regard, this paper proposes an adaptive scheme in a new distributed computing framework that involves HAPS computing to deal with the two problems of the IoV. Based on the diverse demands of vehicles, network dynamics, and the time-sensitivity of handoffs, the proposed scheme flexibly divides each task into three parts and assigns them to the vehicle, roadside units (RSUs), and a HAPS to perform synchronous computing. The scheme also constrains the computing of tasks at RSUs such that they are completed before handoffs to avoid the risk of computing interruptions. On this basis, we formulate a delay minimization problem that considers task-splitting ratio, transmit power, bandwidth allocation, and computing resource allocation. To solve the problem, variable replacement and successive convex approximation--based method are proposed. The simulation results show that this scheme not only avoids the negative effects caused by handoffs in a flexible manner, it also takes delay performance into account and maintains the delay stability.  
\end{abstract}


\section{Introduction}
\subsection{Background}
With advances in technologies such as wireless communication, the Internet of things, and artificial intelligence (AI), a concept of Internet of vehicles (IoV) has also gained great momentum, where autonomous driving is considered to be the ultimate goal \cite{vcompu1,vehicle4}. An intelligent and connected vehicle (ICV), which is the main component of the IoV, will integrate sensing, decision-making, and control functions. In the future, ICVs will realize the transformation of massive information from physical space to information space, and this transformation will depend on fine-grained data computation, including various machine learning–based models of training and inference
 \cite{mingzhechen1,vehicle2}. In the past few years, the field of machine learning has undergone a major shift from a big-data paradigm of centralized cloud processing to a small-data paradigm of distributed processing by devices at a network edge\cite{jsac_chenmingzhe,editor3}. The impetus for this shift is to synchronously utilize locally available resources and nearby resources of edge nodes to obtain real-time responses for AI-based task computing, thereby facilitating the development of distributed computing \cite{reviewer4,mingzhechen2}. Due to the advantages of distributed computing, such as proximity, efficiency, scalability, and easy collaboration, it is expected to play an important role in realizing autonomous driving.  \par

Autonomous driving is expected to produce terabytes of data everyday\cite{Car}.
In this context, the IoV of the future is likely to face a continuous shortage of computing resources, and this shortage will be more severe than for ordinary user networks.
This prediction is driven by two reasons: First, unlike ordinary users, which only occupy resources when they are in use, ICVs are generally online for a long time, and therefore they occupy computing resources continuously. Secondly, IoV services require a higher level of security than ordinary user services. Providing good IoV services is directly related to road and personal safety, which means that a shortage of computing resources will present serious security problems \cite{vehicle3}. To address this problem, a variety of solutions have been proposed, including vehicle collaborative computing,  unmanned aerial vehicle (UAV) computing and satellite computing to complement edge computing and form new distributed computing frameworks\cite{V2V,UAV2V,Sate_IOV,editor1}. However, interruptions between vehicle-to-vehicle connections, short and unreliable UAV dwelling times,  prohibitive satellite transmission delays, and complicated mobility management make it difficult to meet the demands of IoV in terms of security, stability, and reliability. This motivates us to explore other, more feasible solutions. More recently, high altitude platform station (HAPS) systems have been proposed as candidates
for 6G networks \cite{HAPcomputing4, reviewer2}. A HAPS is aerial platform, such as an airship or balloon, capable of long-term deployment as a wireless communication station in the stratosphere, where it has a bird's-eye line-of-sight (LoS) view over a large ground area (with a radius of 50-500 km)\cite{ITU}.  Since HAPSs have large payloads (usually $\ge$ 100 kg), they can carry a variety of resources (antennas, capacity, computation, storage, and so on), which can play a powerful supplementary role for ground computing systems\cite{reviewer1,WaelITS}. Besides, current and future energy conversion techniques for solar energy and wind power as well as battery techniques can provide HAPS with a powerful energy supply potential\cite{wind,battery,Gunes1}.  Much work has been done to demonstrate the gains for introducing HAPS computing to IoT networks in terms of energy and delay performances \cite{HAPcomputing, Gunes2}.  Inspired by this, we introduced HAPS computing into vehicular networks in our recent work \cite{QiqiITS} and proposed a computation offloading scheme to accelerate IoV services.

In addition to the prospect of severe shortages of computing resources, the IoV of the future also faces the issue of network handoffs. In practice,  edge nodes usually have a small communication coverage (less than 300 m \cite{ShuaiyuSAGIN}). To ensure the connections of networks and the sustainability of services, ICVs will inevitably trigger the network connection handoff when they cross the coverage of different edge nodes. 
However, when a network handoff occurs, the ICV may experience a temporary interruption of the offloading process, or it may be unable to directly receive computation results due to being beyond the coverage of the computing target node. To address this, researchers have proposed a series of computing offloading solutions to cope with handoff issues \cite{hand40,hand41,hand43,hand44,hand45,hand46}.  The authors in \cite{hand40,hand41} considered computational performance metrics to design handoff schemes. In  \cite{hand40}, they studied  how to maximize the connection time between vehicles and edge servers while minimizing the number of handoff times. In \cite{hand41}, while taking a shortage of edge server resources into consideration, they used remote cloud computing  resources to assist in computing the tasks that could not be processed in time by the connected edge servers on the basis of jointly optimized network handoff and task migration decisions. However, the issue of IoV handoffs can be affected by a variety of factors, including received signal strength, vehicle speed, reliability, efficiency. Moreover, considering multi-dimensional factors in making handoff decisions complicates the problem-solving process \cite{JsacHand,2007An,JSAC_hand2}. In this regard, \cite{hand43,hand44,hand45,hand46} designed computing migration schemes to adapt to network handoffs, and eliminate their negative impact on computing performance. Accordingly, \cite{hand43}  proposed a scheme to migrate tasks among multiple servers when a handoff occurred. \cite{hand44} set a ``hold on" mode for tasks, where vehicles would suspend task scheduling when tasks could not be processed before connecting to the next available server. Given that task results might not reach a vehicle due a handoff, H. Zhang et al. proposed to migrate the results among multiple servers \cite{hand45}. Based on the assumption that path planning is predictable, \cite{hand46} proposed a predictive uploading and downloading of computing tasks and results to improve the efficiency of result acquisition.  

\subsection{Motivation and contributions}

From existing solutions, we know that the dominant approaches for adapting to edge network handoffs in distributed computing involves computing migration among edge nodes. However, these approaches are inflexible and their performance cannot be guaranteed. Specifically, most approaches have proposed to take compensatory measures for the computing tasks affected by handoffs, such as task and result migration, task re-offloading, and task postponement. While these designs aim to ensure that tasks can be performed or results successfully delivered, such approaches are passive and inflexible. Moreover, since only edge node resources are considered, tasks or their results usually need to be migrated among multiple edge nodes, which will cause a long response time or even unresponsiveness, thus resulting in a wildly fluctuating performance or even computing failure. Therefore, to cope with handoffs, the computing offloading scheme needs to avoid the negative impact of handoff on computing, and at the same time it needs to ensure the low-latency performance of computing tasks in both handoff and non-handoff cases to provide a smooth and stable service experience for ICVs.
  
   Fortunately, a HAPS can help solve the above mentioned problems. With its bird's-eye view of the ground and powerful computing resources, a HAPS can help absorb atypical demand surges and avoid computing interruptions caused by edge network handoffs. Therefore, the introduction of HAPS computing into IoV systems can both alleviate computing resource shortages and effectively deal with the handoff issue, which makes perfect sense for ICVs. In view of this critical observation, we follow the distributed computing framework proposed in our recent work \cite{QiqiITS}. On this basis, we propose a handoff-aware computing offloading scheme to deal with the computing interruption problem. In this scheme, the task will be dynamically split and distributedly assigned to the ICV, RSU, and HAPS nodes for synchronous processing, which means that an ICV's task can be processed in parallel through the coordination of these three node types. Accordingly, when a network handover is about to happen, the RSU is only responsible for a very small amount of data processing to ensure that the task portion assigned to this node can be computed before the handoff occurs, while the HAPS bears most of the computational burden to smoothly complete the computing of the task.
   The contributions of this work are as follows:
   \begin{enumerate}
   	\item  We develop a handoff-aware computing scheme by exploiting a distributed computing framework that integrates the computing resources of ICVs, RSUs, and HAPS nodes.  Different from the existing solutions in edge network, this scheme utilizes the HAPS-assisted distributed framework to deal with the computing interruption issue caused by edge network handoffs in a flexible manner. To the best of our knowledge, this is the first work to investigate the benefits of HAPS computing for this topic.\footnote{In this work, we follow the integrated three-layer computing framework in the previous work \cite{QiqiITS}, and also propose a new computing offloading scheme to handle the computing interruption issue caused by edge network handoff.}
   	\item  Due to the diverse needs of ICVs, network dynamics, and the time-sensitivity of handoffs, this scheme flexibly divides the tasks of each user into three parts and assigns them to ICVs, RSUs, and HAPS nodes to perform synchronous computing. In addition, this scheme constrains the computing of tasks at the RSUs so that they are completed before a handoff is about to occur, thereby avoiding the risk of a computing interruption.
   	\item With the objective of minimizing the delay in executing ICVs' tasks, we formulate an optimization problem by finding task-splitting ratios, ICV transmit power, bandwidth allocation, and computing resource allocation. Then, using variable replacement and successive convex approximation (SCA) methods, we solve the optimization problem.
   \end{enumerate} 

\begin{figure*}[!t]
	\centerline{\includegraphics[height=11cm]{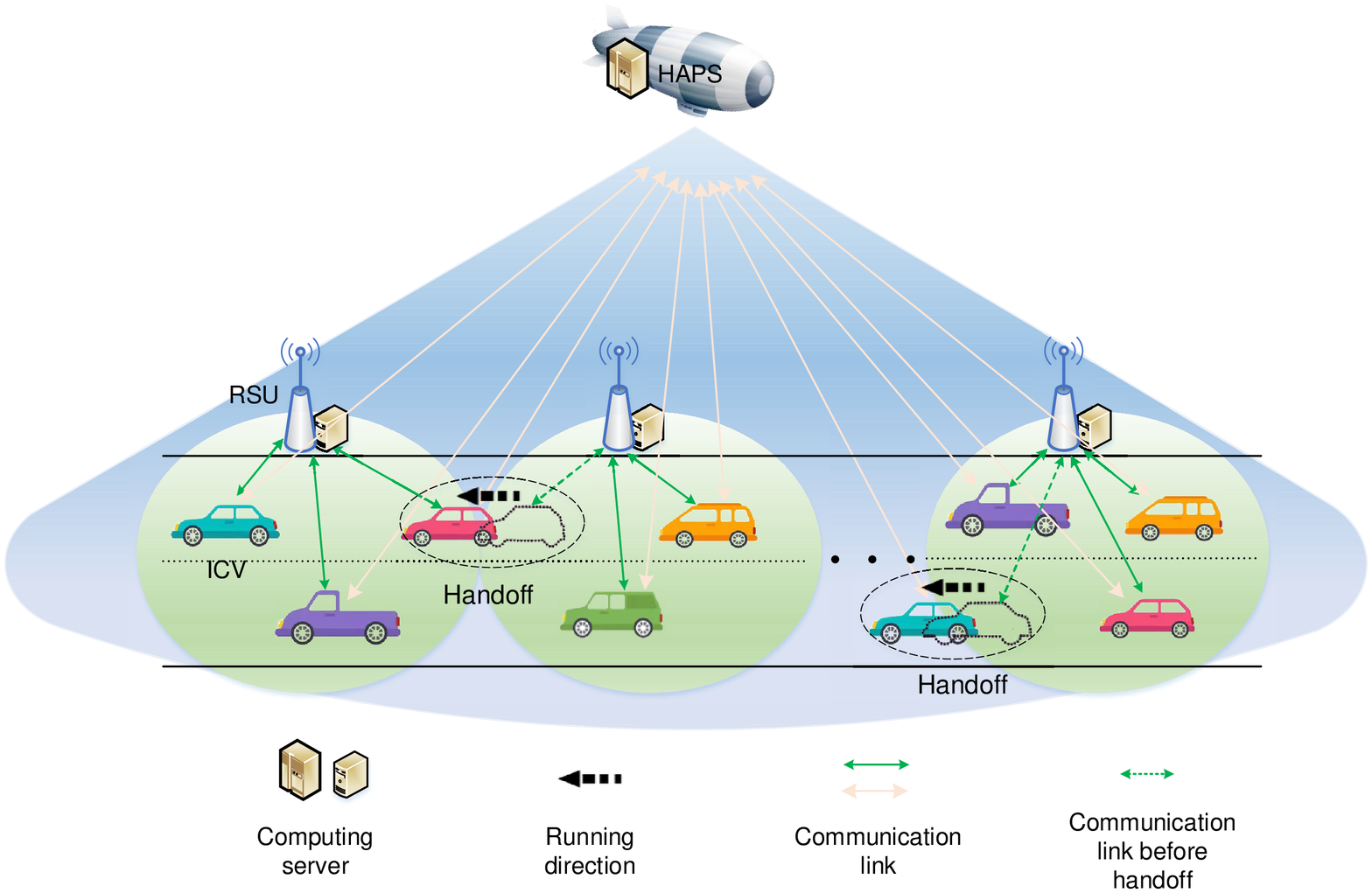}}
	\caption{Handoff-aware distributed computing in in a HAPS-assisted vehicular network.}
	\label{fig1}
\end{figure*}  
   
\section{Handoff-Aware Distributed Computing Model}
\subsection{System Model}
		The handoff-aware distributed computing model considered here for a HAPS-assisted vehicular network is shown in Fig. 1. A single HAPS at an altitude of 20 km provides coverage for a one-way road \cite{TWC1}. The road is divided into $M$ non-overlapping segments, each covered by an edge network access point, i.e., RSU. Each RSU is equipped with a computing server to process vehicular computing tasks, and both RSUs and their equipped computing resources can be considered as edge network resources in the system. These RSUs are labeled as $\mathcal{M}=\{1,2,\dots,M\}$ in order.  We assume that the coverage range of the RSUs is $D $ (m). There are $N$ ICVs, labeled as  $\mathcal{N}=\{1,2,\dots,N\}$, running at a speed of $v$ (m/s) along the road, and each ICV passes through the coverage range of the RSUs in order. This means that a handoff occurs when an ICV passes from one RSU coverage area to another.   At the beginning of each decision time slot, the ICV $n$ runs from its current position $l_n$ (m), and generates one task to be computed, with the input data size $\varepsilon_n$ (in bits) and computational density $\lambda_n$ (in CPU Cycle/bit). Assuming the task is splittable, we consider a distributed parallel computing model, which is shown in Fig. 2. In this model, the task can be split into three portions and computed at three nodes simultaneously (i.e., at the ICV's onboard device, at the associated RSU, and at the HAPS), which corresponds to three computing ways: local, RSU, and HAPS computing, respectively. Local computing can directly process the task portion it is assigned, while the task portions that are assigned to the RSU and HAPS need to be offloaded to the RSU and HAPS separately before the computing can be performed on their servers. 
		In this work, we do not discuss the downloading of results because the amount of such data is small compared to the input data.  Given that delay performance is critical for IoV services, and this performance influences vehicle and road safety as well as the driver's experience, this work is dedicated to improving the delay performance of ICVs in performing computing tasks.  
		\begin{figure*}[!t]
			\centerline{\includegraphics[height=6.8cm]{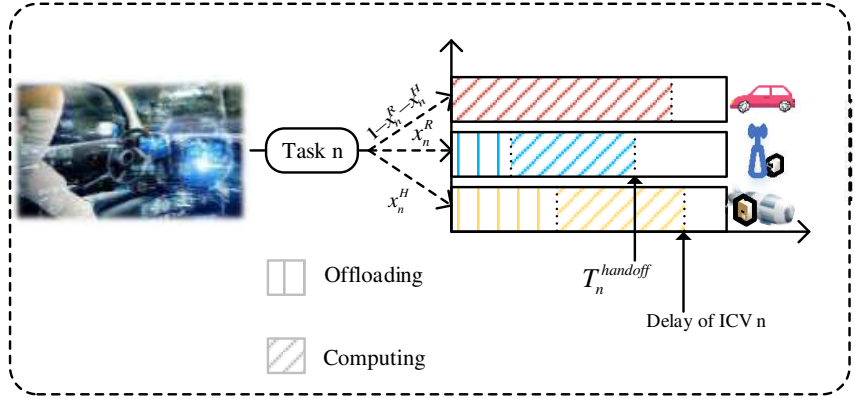}}
			\caption{Distributed parallel computing model.}
			\label{fig2}
		\end{figure*}  
	
		Based on the model shown in Fig. 2, the data splitting model is described as follows: for ICV $n$, the data splitting ratio at its associated RSU node is indicated by $x_n^R$, at the HAPS node is indicated by $x_n^H$, and at its onboard device is indicated by $1-x_n^R-x_n^H$.  The task splitting strategy is dynamically determined rather than using a fixed assignment. It depends on real-time channel conditions, the association between users and RSUs, variational task requirements (data volume and computational density), and available bandwidth and computing resource capacity \cite{editor2}. In addition to these factors, the handoff issue of the edge network is also considered by adjusting the data splitting ratios in a flexible manner so that the task portion computed at the RSU is not interrupted by a handoff. Fig. 3 provides an illustration of a handoff. As we can see, at the beginning of a decision time slot, the ICV $n$ moves from position $A$ to network boundary $B$ and is about to enter the coverage of the next RSU, which triggers a network handoff. According to the ICV's initial speed $v$ (m/s) and its position $l_n$, the handoff timestamp can be given by 
		$T_n^{handoff}=\frac{D-|l_n\mod D|}{v}$,
		where $|l_n\mod D|$ is the remainder of dividing $l_n$ by $D$, which indicates ICV $n$'s relative position from its associated RSU\footnote{The model is formulated according to the  distance-based handoff protocol rules.} $T_n^{handoff}$ describes the time it takes for ICV $n$ to run out of the RSU coverage boundary. During this journey,  ICV $n$ offloads a partial task with a splitting ratio $x_n^R$ to the RSU, and then the RSU server computes the assigned task portion.  To ensure that the data processing will not be interrupted by the network handoff, the delay experienced must be guaranteed not to exceed the timestamp when the handoff occurs, i.e., $T_n^{handoff}$, which can be viewed as a hard time limit. If the processing delay of the task portion at RSU is greater than $T_n^{handoff}$, the corresponding portion will be dropped.

		\begin{figure}[!t]
			\centerline{\includegraphics[height=6.5cm]{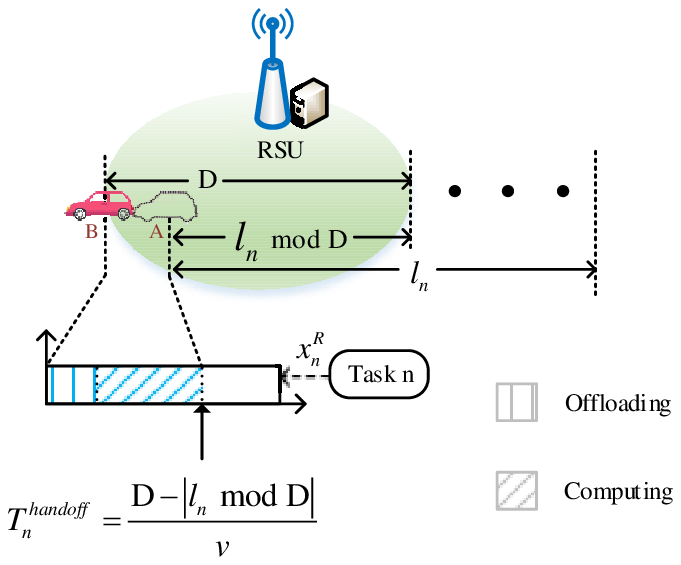}}
			\caption{Illustration of handoff.}
			\label{fig3}
		\end{figure}
\subsection{Communication Model}
In this system, we consider that both RSUs and ICVs are equipped with a single antenna, while the HAPS is equipped with multiple antennas. Based on this assumption, there are a very high number of spotbeams in HAPS network and that the target area is within a spotbeam\cite{QiqiITS}.   At each time slot, the ICV $n$ needs to offload the input data of the task portion with size $x_n^R\varepsilon_n$ to the associated RSU, while offloading the input data of the task portion with size $x_n^H\varepsilon_n$ to the HAPS at the same time. Therefore, the communication links are of two types: links between ICVs and RSUs; and links between ICVs and the HAPS.  We consider a carrier frequency of 2 GHz for all links, and these links work on orthogonal channels. Given the available bandwidth ${B_{\max}}$, the bandwidths for these orthogonal links are optimally allocated. For ICV $n$, the bandwidth ratios allocated to the links from the ICV to the RSU and from the ICV to the HAPS are indicated by $b_n^R$ and $b_n^H$, respectively. The bandwidth allocation should satisfy the compacity constraint $\sum\limits_{n\in\mathcal{N}}b_n^R+b_n^H\le 1$. In addition, due to the limitations of the transmit power of ICVs, the power allocation is also investigated with the given $P_{\max}$. For ICV $n$, the power allocation ratios for the links from ICV $n$ to the RSU and from ICV $n$ to the HAPS are indicated by $p_n^R$ and $p_n^H$, respectively. Here, we use $p_n^R+p_n^H\le 1, \forall n\in\mathcal{N}$ to constrain the  power allocation for the ICVs.  We consider non-line-of-sight (NLoS) communication for the  link from ICV $n$ to the RSU, and the channel gain is modeled as
\begin{equation}\label{op2}
	\begin{aligned}
		G_n^{R} = \frac{{{\beta _0}({f_c}){{\left| {h_n^{R}} \right|}^2}}}{{{{\left( {{d_n^R}} \right)}^\alpha }}},
	\end{aligned}	
\end{equation}
where $\beta_0({f_c})$ is the path loss at the reference distance 1 m,  $h_n^{R}$ is the small-scale  fading coefficient of the NLoS link following a Rayleigh distribution, $d_n^R$ is the  distance from ICV $n$ to the RSU, and $\alpha$ is the path-loss exponent. The corresponding transmission rate  is given by $\small{R_n^{R} = b_n^{R}B_{\max}{\log _2}\left(1 + \frac{{{p_n^R}P_{\max}G_n^{R}}}{{{b_n^{R}}B_{\max}{N_0}}}\right)}$, where  $N_0$ is the Gaussian noise power
spectrum density.   

 Since the HAPS hovers in a high altitude and the highway is usually in a remote area, we consider the links from the ICVs to the HAPS to be LoS with their large-scale fading path loss following free-space path loss. The channel gain can be modeled by \cite{TWC2}:
\begin{equation}\label{op1}
	\begin{aligned}
		G_n^H = {G}{\left( {\frac{c}{{4\pi {d_n^H}{f_c}}}} \right)^2}{\left| {{h_n^H}} \right|^2},
	\end{aligned}
\end{equation}
where  $c$ is the speed of light, $d_n^H$ is the  distance from ICV $n$ to the HAPS. $f_c$ is the carrier frequency, which in this work we consider to be 2 GHz. Since environmental effects are negligible for the frequencies under 10 GHz, environmental attenuations are not considered in this work\cite{HAPsurvey2005}.  $G$ is the directional antenna gain, and $h_n^H$ is the small-scale fading coefficient  corresponding to Rice fading that considers the LoS component. The corresponding transmission rate  is given by $\small{R_n^{H} = b_n^{H}B_{\max}{\log _2}\left(1 + \frac{{{p_n^H}P_{\max}G_n^{H}}}{{{b_n^{H}}B_{\max}{N_0}}}\right)}$.
With the transmission rate, the  delay for offloading the task portion of ICV $n$ with size $x_n^R\varepsilon_n$ to the RSU can be modeled as $\frac{x_n^R\varepsilon_n}{R_n^R}$, and the delay for offloading the task portion of ICV $n$ with size $x_n^H\varepsilon_n$  can be modeled as $\frac{x_n^H\varepsilon_n}{R_n^H}$. In addition, the propagation delays taken for the ICV's signals to reach the RSU and HAPS are given by $\frac{d_n^R}{c}$ and $\frac{d_n^H}{c}$. 


\begin{table*}\nonumber
	\caption*{TABLE I: Notation Definitions}
	
	\begin{center}
		\begin{tabular}{c|c}		
			\hline
			{Notation} & {Definition}  \\ 
			\hline
			{$\mathcal{M}, M$} & {The RSU set and the number of RSUs}  \\ 
			\hline
			{$\mathcal{N}, N$}& {The ICV set and the number of ICVs } \\ 
			\hline
			{$\varepsilon $} & {Volume of input data (Kbits)} \\ 
			\hline
			{$\lambda$} &{ Computation density (CPU cycle/bit)}  \\ 
			\hline
			{$x_n^R,x_n^H$} & {The data splitting ratio at RSU node, and HAPS node} \\ 
			\hline		
			{$D $ } &{ The coverage range of RSU (m)} \\ 
			\hline
			{$F^L,F^R,F^H$} &{ Computational capability of ICV, RSU and HAPS (CPU cycle/s)}  \\ 
			\hline
			{$G^{R}, G^{H}, G$} & {Link channel gain of RSU,  link channel gain of HAPS, and directional antenna gain} \\ 
			\hline
			{$d^{R}, d^{H}$} &{ The link distance of RSU and HAPS (m)} \\ 
			\hline
			{$c, f_c,\alpha, \beta_0$ }&	{Light speed (m/s), carrier frequency (Hz),  NLoS link path loss factor, and NLoS link reference path loss} \\ 
			\hline
			{$ h^{R}, h^{H}$} &{LoS link small-scale fading coefficient  and NLoS link small-scale fading coefficient}  \\ 
			\hline
			{$N_0$} & {Gaussian noise power
				spectrum density (dBm/Hz)}  \\ 
			\hline
			{$B_{\max}, R^R, R^H$}	 & {Bandwidth limitation (MHz),  transmission rate from ICV to RSU,  and transmission rate from ICV to HAPS (bit/s) }\\ 
			\hline
			{$f, b$} &{ Computing resource allocation ratio and bandwidth allocation ratio} \\ 
			\hline
			{$P_{\max}$ }& { Transmitter power limitation of ICV (dBm) }  \\
			\hline
			{$T^L, T^R, T^H$} & {Delay under local, RSU, and HAPS computing (s) }  \\
			\hline
			{$T_n^{handoff}$} & {The time to triger handoff 
			(s) }  \\
			\hline
		\end{tabular}
	\end{center}
\end{table*} 
\subsection{Computing Model}
It is assumed that at each decision time slot, each ICV generates one task to compute that can be split into three portions and computed at three nodes in parallel. The available computing resources for ICVs doing local, RSU, and HAPS computing are indicated by $F^{L}$, $F^{R}$, and $F^{H}$ (in CPU Cycle/s), respectively.  After the splitting, the task portion assigned to the RSU with workloads $x_n^R\varepsilon_n\lambda_n$ will be computed at the RSU server, the task portion assigned to the HAPS with workloads $x_n^H\varepsilon_n\lambda_n$ will be computed at the HAPS, and the task portion assigned to the ICV with workloads $(1-x_n^R-x_n^H)\varepsilon_n\lambda_n$ will be computed locally. The computing resources will be optimized at the RSU servers and HAPS server. For RSU computing, the computing resources will be allocated among the ICVs in the RSU's covered segment. We let $\Phi_m$ denote the ICV index set of RSU $m$, $m\in\mathcal{M}$.  Let $f_n^R$ indicate the computing resource ratio allocated for ICV $n$ at its associated RSU server, so the computing resource allocation limitation at RSU $m$ is given by $\sum\limits_{n\in\Phi_m} f_n^R\le1,\forall m\in \mathcal{M}$. Let $f_n^H$ indicate the computing resource ratio allocated at the HAPS server for ICV $n$, and the computing resource allocation limitation at HAPS is constrained by $\sum\limits_{n\in\mathcal{N}}f_n^H\le1$.  Furthermore,  the computational delays can be expressed as $\frac{(1-x_n^R-x_n^H)\varepsilon_n\lambda_n}{F^{L}}$ for local computing,  $\frac{x_n^R\varepsilon_n\lambda_n}{f_n^RF^{R}}$ for RSU computing, and $\frac{x_n^H\varepsilon_n\lambda_n}{f_n^HF^{H}}$ for HAPS computing, respectively. 
 
\subsection{Delay Model} 
Using the above communication and computing models, the delays for ICV $n$ with local, RSU, and HAPS computing can be given as follows: 

\noindent 1) Local computing:
	\begin{equation}\label{op6}
		T_{n}^L= \dfrac{{(1 - x_n^R - x_n^H){\varepsilon _n}{\lambda _n}}}{{{F^L}}}.
	\end{equation}
2) RSU computing:
	\begin{equation}\label{op7}
	T_n^R=  \frac{d_n^R}{c}+{\frac{{x_n^R{\varepsilon _n}}}{{b_n^R{B_{\max }}{{\log }_2}\left( {1 + \frac{{p_n^R{P_{\max }}G_n^R}}{{b_n^R{B_{\max }{N_0}}}}} \right)}}} + \frac{{x_n^R{\varepsilon _n}{\lambda _n}}}{{f_n^R{F^R}}}.
	\end{equation}
3) HAPS computing:
	\begin{equation}\label{op6}
		T_n^H= \frac{d_n^H}{c}+ {\frac{{x_n^H{\varepsilon _n}}}{{b_n^H{B_{\max }}{{\log }_2}\left( {1 + \frac{{p_n^H{P_{\max }}G_n^H}}{{b_n^H{B_{\max }{N_0}}}}} \right)}}} + \frac{{x_n^H{\varepsilon _n}{\lambda _n}}}{{f_n^H{F^H}}}.
	\end{equation}
The delay for ICV $n$ is ultimately determined by the maximum value of $T_n^L, T_n^R$ and $T_n^H$ \footnote{ According to  $T_c=\lambda/v$, in which $\lambda$ is wavelength and $v$ is vehicle velocity, the coherence time in our scenario is 5 milliseconds. During the coherence time, only 66.6 microseconds is assigned for the channel estimation (under the assumption of a 15 KHz subcarrier, and allocation of one symbol per each pilot signal), and the 
		remaining time can be allocated for offloading.
		This means that the channel estimation time is negligible compared to the offloading time.  }.
\section{Sum-Delay Minimization}
In this section, we first formulate the optimization problem that minimizes the total delay of all ICVs in order to improve the average delay performance of the system. Then, we provide the problem transformation and solution for the formulated problem. 
\subsection{Problem Formulation}
We aim to optimize the sum-delay experienced by ICVs for executing tasks by finding the optimal values for the task-splitting ratios of RSU $\bm{{\rm X}^R}{ = \text{[}x}_1^R,...,{x}_N^R{\text{]}}$, the task-splitting ratios of HAPS $\bm{{\rm X}^H}{ = \text{[}x}_1^H,...,{x}_N^H{\text{]}}$, bandwidth allocations of the links from ICVs to RSU $\bm{{\rm B}^R}{ = \text{[}b}_1^R,...,{b}_N^R{\text{]}}$, bandwidth allocations of the links from ICVs to HAPS $\bm{{\rm B}^H}{ = \text{[}b}_1^H,...,{b}_N^H{\text{]}}$, power allocations of the links from ICVs to RSU $\bm{{\rm P}^R}{ = \text{[}p}_1^R,...,{p}_N^R{\text{]}}$, power allocations of the links from ICVs to HAPS $\bm{{\rm P}^H}{ = \text{[}p}_1^H,...,{p}_N^H{\text{]}}$, computing resource allocations of RSU $\bm{{\rm F}^R}{ = \text{[}f}_1^R,...,{f}_N^R{\text{]}}$, and computing resource allocations of HAPS $\bm{{\rm F}^H}{ = \text{[}f}_1^H,...,{f}_N^H{\text{]}}$. The optimization problem can be formulated as follows:
\begin{subequations}\label{op7}
	\begin{align}
		&\mathop {\min }\limits_{\begin{subarray}{l}
				\begin{gathered}
					\bm{x^R},\bm{x^H},
					\bm{b^R},\bm{b^H},\hfill \\
					\bm{p^R},\bm{p^H},
					\bm{f^R},\bm{f^H} \\			
				\end{gathered}
		\end{subarray}}~~~    { \sum\limits_n^N {{\max~~\left\{T_n^L,T_n^R,T_n^H\right\}}} }\label{op7a}\\
		\text{s.t.}~~~
		&T_n^R \le T_n^{handoff},~~\forall{n}\in\mathcal{N},\label{op7b}\\
		&\sum\limits_{n\in\mathcal{N}} {{b_n^R + b_n^H}}  \le 1,\label{op7c}\\
		&p_n^R + p_n^H \le 1,~\forall{n}\in\mathcal{N},\label{op7d}\\
		&\sum\limits_{n \in {\Phi _m}} {f_n^R}  \le 1,~\forall{m}\in\mathcal{M},\label{op7e}\\
		&\sum\limits_{n\in\mathcal{N}} {f_n^H}  \le 1,\label{op7f}\\
		&x^R_n,x^H_n,b^R_n,b^H_n,p^R_n,p^H_n,f^R_n,f^H_n>0\label{op7g}.
	\end{align}
\end{subequations}
Since we consider parallel task computing, the ICV's delay is the maximum value of the delay to perform the task on the local, RSU, and HAPS, i.e., $\max~\left\{T_n^L,T_n^R,T_n^H\right\}$. 
In Problem \eqref{op7}, the objective function \eqref{op7a} is the total delay of ICVs. 
   The constraint \eqref{op7b} indicates that the task portion that is assigned to the RSU should be completed before the handoff occurs. \eqref{op7c} denotes the bandwidth constraint for all links. \eqref{op7d} denotes the transmit power constraint for each ICV. \eqref{op7e} denotes the computing resource constraint for each RSU, and \eqref{op7f} denotes the computing resource constraint for HAPS. Finally, \eqref{op7g} indicates the non-negative requirements for all variables. 
It is clear that Problem \eqref{op7} is nonconvex due to the non-smooth objective function, coupled variables, and complicated formulations. This is a difficult problem to solve directly, and there is no standard method to address this problem. In the following, we transform the problem into a tractable one.  
\subsection{Problem Transformation and Solution} 
 First of all, we introduce auxiliary variable $\bm{\textrm{T}}{ = \text{[}T}_1,...,{T}_N{\text{]}}$ to represent the individual delays, so that the original non-smooth problem is transformed into a smooth one. Further, by expanding the representation of $T_n^L, T_n^R$, and $T_n^H$, Problem \eqref{op7} can be equivalently rewritten as
\begin{subequations}\label{op8}
	\begin{align}
		&\mathop {\min }\limits_{\begin{subarray}{l}
				\begin{gathered}
					\bm{\textrm{T}},\bm{x^R},\bm{x^H},
					\bm{b^R},\bm{b^H},\hfill \\
					\bm{p^R},\bm{p^H},
					\bm{f^R},\bm{f^H} \\			
				\end{gathered}
		\end{subarray}}~~~    { \sum\limits_n^N {~~~{T_n}} }\label{op8a}\\
		\text{s.t.}~~~
		&\dfrac{{(1 - x_n^R - x_n^H){\varepsilon _n}{\lambda _n}}}{{{F^L}}}\le T_n,\  \forall n \in\mathcal{N},\label{op8TL}\\
		&\frac{d_n^R}{c}+ {\frac{{x_n^R{\varepsilon _n}}}{{b_n^R{B_{\max }}{{\log }_2}\left( {1 + \frac{{p_n^R{P_{\max }}G_n^R}}{{b_n^R{B_{\max }{N_0}}}}} \right)}}} + \frac{{x_n^R{\varepsilon _n}{\lambda _n}}}{{f_n^R{F^R}}}\le T_n, \nonumber \\
		& \quad\quad\quad\quad\quad\quad\quad\quad\quad\quad\quad\quad\quad\quad\quad\quad\forall n \in \mathcal{N},\label{op8TR}\\
		&\frac{d_n^H}{c}+ {\frac{{x_n^H{\varepsilon _n}}}{{b_n^H{B_{\max }}{{\log }_2}\left( {1 + \frac{{p_n^H{P_{\max }}G_n^H}}{{b_n^H{B_{\max }{N_0}}}}} \right)}}} + \frac{{x_n^H{\varepsilon _n}{\lambda _n}}}{{f_n^H{F^H}}}\le T_n,\nonumber \\
		& \quad\quad\quad\quad\quad\quad\quad\quad\quad\quad\quad\quad\quad\quad\quad\quad\forall n \in \mathcal{N},\label{op8TH}\\
		& \frac{d_n^R}{c}+{\frac{{x_n^R{\varepsilon _n}}}{{b_n^R{B_{\max }}{{\log }_2}\left( {1 + \frac{{p_n^R{P_{\max }}G_n^R}}{{b_n^R{B_{\max }{N_0}}}}} \right)}}} + \frac{{x_n^R{\varepsilon _n}{\lambda _n}}}{{f_n^R{F^R}}} \nonumber \\
		& \quad\quad\quad\quad\quad\quad\quad\quad\quad\le T_n^{handoff},~~\forall{n}\in\mathcal{N},\label{op8b}\\
		&\eqref{op7c},\eqref{op7d},\eqref{op7e},\eqref{op7f},\eqref{op7g}.
	\end{align}
\end{subequations}

Next, in order to handle the coupling between the optimization variables in the constraints \eqref{op8TR}, \eqref{op8TH}, and \eqref{op8b}, we introduce auxiliary variables $\bm{\tau^R}{ = \text{[}\tau}_1^R,...,{\tau}_N^R{\text{]}}$ and $\bm{\tau^H}{ = \text{[}\tau}_1^H,...,{\tau}_N^H{\text{]}}$ to respectively represent the delays of ICVs offloading data to the RSU and HAPS, and relax the corresponding constraints. Then, Problem \eqref{op8} can be equivalently rewritten as 
\begin{subequations}\label{op9}
	\begin{align}
		&\mathop {\min }\limits_{\begin{subarray}{l}
				\begin{gathered}
					\bm{\textrm{T}}, \bm{x^R},\bm{x^H},
					\bm{b^R},\bm{b^H},\hfill \\
					\bm{p^R},\bm{p^H},\bm{\tau^R},\bm{\tau^H},
					\bm{f^R},\bm{f^H} \\
				\end{gathered}
		\end{subarray}}~~~\sum\limits_{\text{n}}^N  {T_n}\label{op9a}\\
		\text{s.t.}~~~& \frac{d_n^R}{c}+\tau_n^R + \frac{{x_n^R{\varepsilon _n}{\lambda _n}}}{{f_n^R{F^R}}}\le T_n,~\forall n \in \mathcal{N},\label{op9TR}\\
		& \frac{d_n^H}{c}+\tau_n^H + \frac{{x_n^H{\varepsilon _n}{\lambda _n}}}{{f_n^H{F^H}}}\le T_n,~\forall n \in \mathcal{N},\label{op9TH}\\
		&{\frac{{x_n^R{\varepsilon _n}}}{{b_n^R{B_{\max }}{{\log }_2}\left( {1 + \frac{{p_n^R{P_{\max }}G_n^R}}{{b_n^R{B_{\max }{N_0}}}}} \right)}}} \le \tau _n^R,~\forall{n}\in\mathcal{N},\label{op9b}\\
		&{\frac{{x_n^H{\varepsilon _n}}}{{b_n^H{B_{\max }}{{\log }_2}\left( {1 + \dfrac{{p_n^H{P_{\max }}G_n^H}}{{b_n^H{B_{\max }{N_0}}}}} \right)}}} \le \tau _n^H,~\forall{n}\in\mathcal{N},\label{op9c}\\
		& \frac{d_n^R}{c}+\tau _n^R +  \frac{{x_n^R{\varepsilon _n}{\lambda _n}}}{{f_n^R{F^R}}} \le T_n^{handoff},~\forall{n}\in\mathcal{N},\label{op9d}\\
		&\eqref{op7c}-\eqref{op7f},\eqref{op8TL},\nonumber\\
		&x^R_n,x^H_n,b^R_n,b^H_n,p^R_n,p^H_n,\tau^R_n,\tau^H_n,f^R_n,f^H_n>0,\label{op9e}
	\end{align}
\end{subequations}
where \eqref{op9b} guarantees that the communication delays experienced by the ICVs offloading data to the RSU cannot exceed $\pmb{\tau^R}$, and \eqref{op9c} guarantees that the communication delays experienced by the ICVs offloading data to the HAPS cannot exceed $\pmb{\tau^H}$. Then, to further address the variable coupling issue, we make the exponential transformations for variables $\bm{x^R},\bm{x^H},\bm{\tau^R},\bm{\tau^H},\bm{f^R}$~and~$\bm{f^H}$. More specifically, the above variables for ICV ~$n$~ can be converted as follows: $x_n^R=\exp(\overline{x_n^R}), x_n^H=\exp(\overline{x_n^H}), \tau^R_n=\exp(\overline {\tau _n^R}), \tau^H_n=\exp(\overline {\tau _n^H}), f_n^R=\exp(\overline{f_n^R})$ and $f_n^H=\exp({\overline{f_n^H}})$. Then, we can obtain the following optimization problem: 
\begin{subequations}\label{op10}
	\begin{align}
		&\mathop {\min }\limits_{\begin{subarray}{l}
				\begin{gathered}
					\bm{\textrm{T}},\bm{\overline{x^R}},\bm{\overline{x^H}},
					\bm{b^R},\bm{b^H},\hfill \\
					\bm{p^R},\bm{p^H},
					\bm{\overline{\tau^R}},\bm{\overline{\tau^H}},			
					\bm{\overline{f^R}},\bm{\overline{f^H}} \hfill \\
				\end{gathered}
		\end{subarray}} ~~~   { \sum\limits_n^N {{T_n}} }\label{op10a}\\
		\text{s.t.}~	&\frac{{{\varepsilon _n}{\lambda _n}}}{{{F^L}}}\left(1 - \exp \left(\overline {x_n^R}  \right) - \exp \left(\overline {x_n^H}  \right)\right)\le T_n, \ \forall n,\label{op10TL}\\
		& \frac{d_n^R}{c}+\exp{\left(\overline {\tau_n^R}\right)} + \frac{{{\varepsilon _n}{\lambda _n}}}{{{F^R}}}\exp{\left(\overline {x_n^R}-\overline {f_n^R}\right)}\le T_n,~\forall n \in \mathcal{N},\label{op10TR}\\
		&\frac{d_n^H}{c}+\exp{\left(\overline {\tau_n^H}\right)} + \frac{{{\varepsilon _n}{\lambda _n}}}{{{F^H}}}\exp{\left(\overline {x_n^H}-\overline {f_n^H}\right)}\le T_n,~\forall n \in \mathcal{N},\label{op10TH}\\
		&b_n^{R}B_{\max}{\log _2}\left(1 + \frac{{{p_n^R}P_{\max}G_n^{R}}}{{{b_n^{R}}B_{\max}{N_0}}}\right) \ge {\varepsilon _n}\exp \left(\overline {x_n^R}  - \overline {\tau _n^R} \right),\nonumber \\
		& \quad\quad\quad\quad\quad\quad\quad\quad\quad\quad\quad\quad\quad\quad\quad\quad\forall{n}\in\mathcal{N},\label{op10b}\\
		&b_n^{H}B_{\max}{\log _2}\left(1 + \frac{{{p_n^H}P_{\max}G_n^{H}}}{{{b_n^{H}}B_{\max}{N_0}}}\right) \ge {\varepsilon _n}\exp \left(\overline {x_n^H}  - \overline {\tau _n^H} \right),\nonumber \\
		& \quad\quad\quad\quad\quad\quad\quad\quad\quad\quad\quad\quad\quad\quad\quad\quad\forall{n}\in\mathcal{N},\label{op10c}\\
		&\frac{d_n^R}{c}+ \exp \left(\overline {\tau _n^R} \right) + \frac{{{\varepsilon _n}{\lambda _n}}}{{{F^R}}}\exp \left(\overline {x_n^R}  - \overline {f_n^R} \right) \le T_n^{handoff},\nonumber \\
		& \quad\quad\quad\quad\quad\quad\quad\quad\quad\quad\quad\quad\quad\quad\quad\quad\forall{n}\in\mathcal{N},\label{op10d}\\
		&\sum\limits_{n \in {\Phi _m}} {\exp \left(\overline {f_n^R} \right)}  \le 1,~\forall{m}\in\mathcal{M},\label{op10e}\\
		&\sum\limits_{n\in\mathcal{N}} {\exp \left(\overline {f_n^H} \right)}  \le 1,\label{op10f}\\
		&\overline{x^R_n},\overline{x^H_n},b^R_n,b^H_n,p^R_n,p^H_n,\overline{\tau^R_n},\overline{\tau^H_n},\overline{f^R_n},\overline{f^H_n}>0,\label{op10g}\\
		&\eqref{op7c},\eqref{op7d}.
	\end{align}
\end{subequations}

\noindent\textbf{Proposition 1:} The resulting sets of constraints \eqref{op10TR}-\eqref{op10f} are convex. 
\begin{proof} 
	In the following, the proofs relevant to constraints ~\eqref{op10TR}, \eqref{op10TH}, \eqref{op10d}, \eqref{op10e}, as well as \eqref{op10f} and \eqref{op10b}-\eqref{op10c} are developed separately. 
	\begin{enumerate}		
		\item The left hand of constraints ~\eqref{op10TR}, \eqref{op10TH}, \eqref{op10d}, \eqref{op10e}, and \eqref{op10f} are in the form of a summation of positive terms of exponential functions, so these constraints are all convex.
		\item Constraints \eqref{op10b} and \eqref{op10c} have similar expressions as follows:		
		\begin{equation}\label{proof}
			\theta y{\log _2}(1 + \frac{{\eta z}}{\theta }) - q\exp (\delta  - \omega ) \ge 0,
		\end{equation}
	 where $\alpha, \beta, \delta$, and $\omega$ represent optimization variables, and $y, z,$ and $q$ represent constants. 
	 The first term $e(\alpha, \beta)=\alpha y{\log _2}(1 + \frac{{\beta z}}{\alpha })$ is the perspective form of concave function $\log_2(1+\beta z)$ \cite{cvx}. Accordingly, we can say that $e(\alpha, \beta)$ is jointly concave in variables $\alpha$ and $\beta$. Furthermore, it is easy to prove the second term $-q\exp(\delta-\omega)$ is jointly concave in variables $\delta$ and $\omega$. Therefore, both the resulting sets of constraints \eqref{op10b} and \eqref{op10c} are convex. 
	\end{enumerate}
\end{proof}
According to the proof,  the exponential transformation of variables can effectively remove the coupling between variables. However, Problem \eqref{op10} is still non-convex because of constraint \eqref{op10TL}. To deal with this problem, we use the successive convex approximation (SCA)
method. SCA is widely used for iteratively approximating an originally non-convex problem by using first-order Taylor expansion, which can transform the original problem into a series of convex versions \cite{b4,b5}. In this case, we first define $g_n={1 - \exp (\overline {x_n^R} ) - \exp (\overline {x_n^H} )}  $ for any $n\in\mathcal{N}$. Then, we define $\pmb{\mathcal{A}^i}=\{\mathcal{A}_n^i \big{|} n\in{\mathcal{N}}$\}, wherein $A^i_n= \{\hat{x_n^R}[i],\hat{x_n^H}[i]\}$ is the local point set in the
 $i$-th iteration. Recalling that any concave function
 is globally upper bounded by its first-order Taylor expansion at any given point \cite{Renqiqi2,b50}, we can obtain the following inequality:
 \begin{equation}\label{SCA}
	\begin{aligned}
		\begin{gathered}
			g_n = 1 - \exp (\overline {x_n^R} ) - \exp (\overline {x_n^H} ) \hfill \\
			\le 1 - \exp \left( {\hat{x_n^R}[i] } \right) - \exp \left( {\hat{x_n^R}[i] } \right)\left( {\overline {x_n^R}  - \hat{x_n^R}[i] } \right) \\
			- \exp \left( {\hat{x_n^H}[i] } \right) - \exp \left( {\hat{x_n^H}[i] } \right)\left( {\overline {x_n^H}  - \hat{x_n^H}[i] } \right)= \hat{g_n}[i] \hfill. \\
		\end{gathered}
	\end{aligned}
\end{equation}
By applying the SCA method, the concave term of \eqref{op10TL} is replaced by a series of linear terms with given local points. As a consequence, Problem \eqref{op10} is transformed into a series of iteratively convex problem, and the $i$-th convex problem is given by: 
\begin{subequations}\label{op11}
	\begin{align}
		&\mathop {\min }\limits_{\begin{subarray}{l}
				\begin{gathered}
					\bm{\textrm{T}},\bm{\overline{x^R}},\bm{\overline{x^H}},
					\bm{b^R},\bm{b^H},\hfill \\
					\bm{p^R},\bm{p^H},
					\bm{\overline{\tau^R}},\bm{\overline{\tau^H}},			
					\bm{\overline{f^R}},\bm{\overline{f^H}} \hfill \\
				\end{gathered}
		\end{subarray}} ~~~   { \sum\limits_n^N {{T_n}[i]} }\\
		&\text{s.t.}~~~\frac{{{\varepsilon _n}{\lambda _n}}}{{{F^L}}} \hat{g_n}[i]\le T_n[i],\label{op11TL}\\
		&\eqref{op7c},\eqref{op7d},\eqref{op10TR}-\eqref{op10g}.
	\end{align}
\end{subequations}

The SCA process is described in Algorithm 1 and its computational complexity is presented here. At each iteration,
the computational complexity is determined by solving the convex problem, 
where the convex problem can be solved by using the interior point method. This method requires $\frac{{\log \left( {\frac{{{N_I}}}{{{u^0}\rho }}} \right)}}{{\log \left( \xi  \right)}}$
number of iterations (Newton steps), where $N_I$ is the number of constraints, $u$ is the initial point used to 
estimate the accuracy of the interior-point method, $\gamma $ is the stopping criterion, and $\xi$ is the parameter 
for updating the accuracy of the interior point method \cite{OmidTWC}.  The number of constraints in problem \eqref{op11} is $2+M+4N$,
where $M$ and $N$ represent the number of RSUs and ICVs in the system, respectively. 
Therefore, the computational complexity of Algorithm 1 is $\mathcal{O}\left({i_{\max }\frac{{\log \left( {\frac{{{N_I}}}{{{u^0}\rho }}} \right)}}{{\log \left( \xi  \right)}}}\right)$, where $i_{\max}$ is the 
maximum number of iterations required for Algorithm 1 to converge \footnote{{Note that our  system model of the terrestrial network can be generalized into the multiple antenna system, in which each RSU is equipped with $A$ antennas and serves $N$ singe-antenna ICVs. For this end, the $N$
	ICVs can be divided into $\frac{N}{L}$
	groups in which $L$ is the number of ICVs in each group. Assuming that $A>\frac{N}{L}$,  each RSU can create $\frac{N}{L}$ beams and each of these beams is responsible to serve one group. Considering orthogonal frequency resource blocks are allocated for the $L$ ICVs inside each group, the proposed bandwidth allocation can be applied to each group.}}. 

\begin{algorithm}
	\caption{The SCA-based method for solving Problem \eqref{op10}.}
	\label{alg:A}
	\begin{algorithmic}[1]
		\STATE Initialize precision $\zeta$, $i_{\max}$, $\pmb{\mathcal{A}^0}$, and set $i=0$.
		\REPEAT 
		\STATE According to the given local point set $\pmb{\mathcal{A}^i}$ and the expression \eqref{SCA}, obtain $\pmb{\hat{g}[i]}$.
		\STATE Solve the convex Problem \eqref{op11} and find the optimal solution set $\Lambda  = \{ \pmb{\textrm{T}}^*,\pmb{\overline{x^R}}^*,\pmb{\overline{x^H}}^*,
		\pmb{b^R}^*,\pmb{b^H}^*,
		\pmb{p^R}^*,\pmb{p^H}^*,\pmb{\overline{\tau^R}}^*,\pmb{\overline{\tau^H}}^*,			
		\pmb{\overline{f^R}}^*,\pmb{\overline{f^H}}^*\} $. 
		\STATE Record the objective function value of the $i$-th iteration as $\sum\limits_n^N T_n[i]=\phi^*$, and update $\pmb{\mathcal{A}^{i+1}}$ based on  $\pmb{\hat{x^R}}[i+1]=\pmb{\overline{x^R}}^*$ and $\pmb{\hat{x^H}}[i+1]=\pmb{\overline{x^H}}^*$, then $i=i+1$.  
		\UNTIL  $\phi^*[i] -\phi^*[i+1]\le \zeta$, or $i>i_{\max}$.
	\end{algorithmic}
\end{algorithm}  
\section{{Maximum-Delay Minimization}}
In this section, we discuss a new optimization problem that focuses on the individual delay performance of ICVs by optimizing the delay fairness between them. Indeed, in delay-sensitive applications, we have to consider individual delay performance  by minimizing the maximum-delay value for all ICVs.  Assuming the same parameters as the sum-delay minimization problem, the optimization problem can be expressed as follows:
\begin{subequations}\label{op12}
	\begin{align}
		&\mathop {\min }\limits_{\begin{subarray}{l}
				\begin{gathered}
					\bm{x^R},\bm{x^H},
					\bm{b^R},\bm{b^H},\hfill \\
					\bm{p^R},\bm{p^H},
					\bm{f^R},\bm{f^H} \\			
				\end{gathered}
		\end{subarray}}~~~    { \max\limits_{n\in\mathcal{N}}~~~~{{\max~\left\{T_n^L,T_n^R,T_n^H\right\}}} }\label{op12a}\\
		\text{s.t.}~~~
		&T_n^R \le T_n^{handoff},~~\forall{n}\in\mathcal{N},\label{op12b}\\
		&\sum\limits_{n\in\mathcal{N}} {{b_n^R + b_n^H}}  \le 1,\label{op12c}\\
		&p_n^R + p_n^H \le 1,~\forall{n}\in\mathcal{N},\label{op12d}\\
		&\sum\limits_{n \in {\Phi _m}} {f_n^R}  \le 1,~\forall{m}\in\mathcal{M},\label{op12e}\\
		&\sum\limits_{n\in\mathcal{N}} {f_n^H}  \le 1,\label{op12f}\\
		&x^R_n,x^H_n,b^R_n,b^H_n,p^R_n,p^H_n,f^R_n,f^H_n>0\label{op12g}.
	\end{align}
\end{subequations}
Problem \eqref{op12} minimizes the maximum-delay value of all ICVs by finding the optimal solution of optimization variables. These variables include the task-splitting ratios of RSUs, i.e., $\bm{{\rm X}^R}{ = \text{[}x}_1^R,...,{x}_N^R{\text{]}}$, the task-splitting ratios of HAPS, i.e., $\bm{{\rm X}^H}{ = \text{[}x}_1^H,...,{x}_N^H{\text{]}}$, bandwidth allocations of the links from ICVs to RSUs, i.e., $\bm{{\rm B}^R}{ = \text{[}b}_1^R,...,{b}_N^R{\text{]}}$, bandwidth allocations of the links from ICVs to HAPS, i.e., $\bm{{\rm B}^H}{ = \text{[}b}_1^H,...,{b}_N^H{\text{]}}$, power allocations of the links from ICVs to RSUs, i.e., $\bm{{\rm P}^R}{ = \text{[}p}_1^R,...,{p}_N^R{\text{]}}$, power allocations of the links from ICVs to HAPS, i.e., $\bm{{\rm P}^H}{ = \text{[}p}_1^H,...,{p}_N^H{\text{]}}$, computing resource allocations of RSUs, i.e., $\bm{{\rm F}^R}{ = \text{[}f}_1^R,...,{f}_N^R{\text{]}}$, and computing resource allocations of HAPS, i.e., $\bm{{\rm F}^H}{ = \text{[}f}_1^H,...,{f}_N^H{\text{]}}$. In addition, the constraint \eqref{op12b} indicates that the  delay for computing task on the RSU cannot exceed the network handoff time. Similar to the sum delay minimization problem, constraints \eqref{op12c}-\eqref{op12g} represent the basic constraints, including bandwidth capacity, power threshold, RSU and HAPS computational capabilities, and the non-negative conditions of variables, respectively. Problem \eqref{op12} is clearly a non-convex problem. In order to solve it, similar to the solution of Problem \eqref{op7}, we adopt the following transformation and solution process: First, we introduce auxiliary variables $\bm{\textrm{T}}{ = \text{[}T}_1,...,{T}_N{\text{]}}$ to transform the original objective function as: $\min\ \max\limits_{n\in\mathcal{N}}\ T_n$. Then, we further introduce variable $\mathbb{T}$ to replace $\max\limits_{n\in\mathcal{N}}\ T_n$. Next, in order to deal with the coupling between optimization variables,  the auxiliary variables $\tau_n^R$ and $\tau_n^H$ are introduced for ICV $n\in\mathcal{N}$ to represent the delays for offloading data to the RSU and HAPS, respectively, and the corresponding delay constraints for these newly introduced variables are added. After that, we apply the exponential transformation for the variables  $x_n^R,x_n^H,\tau _n^R,\tau _n^H, f_n^R$ and $f_n^H$. In the transformed optimization problem, the constraint for the local delay will be ~$\frac{{{\varepsilon _n}{\lambda _n}}}{{{F^L}}}\left(1 - \exp \left(\overline {x_n^R} \right) - \exp \left(\overline {x_n^H} \right)\right)\le T_n$,  where the left term is a concave function.  Therefore, the resulting set from this constraint is non-convex. To address this issue, we use the SCA method proposed in Algorithm 1 to transform the original optimization problem into a series of convex sub-problems, and we obtain the final result through the corresponding iterations.

\section{Simulation Results}
For our simulation, we consider a  one-way road covered by two RSUs and one HAPS. The HAPS is deployed in the stratosphere at an altitude of 20 km and located horizontally in the center of the road. All ICVs can communicate with the HAPS instantly and offload their data to the HAPS's server for data processing. Two RSUs were deployed at positions of 80 m and 240 m, each covering a range of 160 m. The ICVs covered by the RSUs can communicate with the RSUs and offload data to their servers for data processing. When the ICV travels beyond the coverage area of the RSU it is connected to, a network handoff will occur. The ICV will then enter the coverage area of the next RSU, establish communication, and obtain services from it. In this simulation, ten ICVs were initially randomly deployed on the road. At the beginning of each decision time slot, each ICV randomly generated one task to be computed, depending on the individual input data. If not emphasized, the main parameters in this simulation were set as in TABLE II.

\begin{table*}\nonumber
	\caption*{TABLE II: Simultation Parameters}
	\begin{center}
		\begin{tabular}{c|c}
			\hline
			Definition & Value \\ 
			\hline
			Speed of ICV (m/s) & 30  \\ 
			\hline
		Bandwidth capacity (MHz) & 20  \\ 
			\hline
			Noise power density (dBm/Hz) & -174 \\
			\hline
			NLoS Path-loss factor & 3.7 \\
			\hline
			Rician factor (dB)  & 10 \\
			\hline
			Rayleigh distribution  & $\mathcal{CN}(0,1)$ \\
			\hline
		Transmitter power of ICV (dBm)  & 23  \\
			\hline
			Directional antenna gain of HAPS (dBi)  & 17 \cite{17dbi}\\
			\hline
			Computational density (CPU cycle/bit) & 500, 1,000, 1,500, 2,000, or 2,500  \\
			\hline 
		Volume of input data (Kbits)  & 100, 300, 500, 700, or 900  \\
			\hline
		Computational capability of\\ ICV, RSU, HAPS (CPU cycle/s)  & 2 G, 32 G, 100 G  \\
			\hline
			Maximum iteration & 50\\
			\hline
			Convergence precision  & $10^{-6}$\\
			\hline
		\end{tabular}
	\end{center}
\end{table*}

\begin{figure}[!t]
	\centerline{\includegraphics[height=6.5cm]{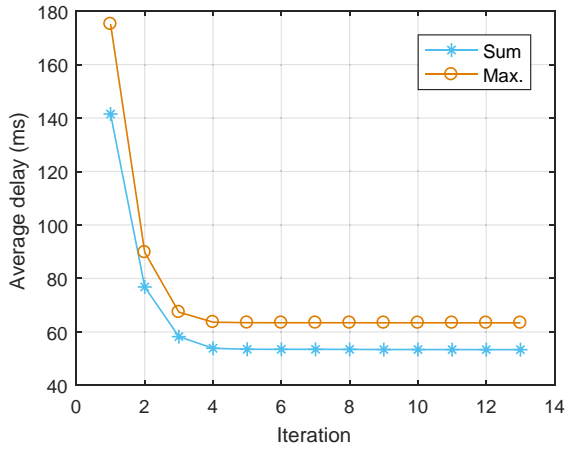}}
	\caption{The convergence of the SCA method with the proposed scheme.}
	\label{fig4}
\end{figure}
\begin{figure}[!t]
	\centerline{\includegraphics[height=10cm]{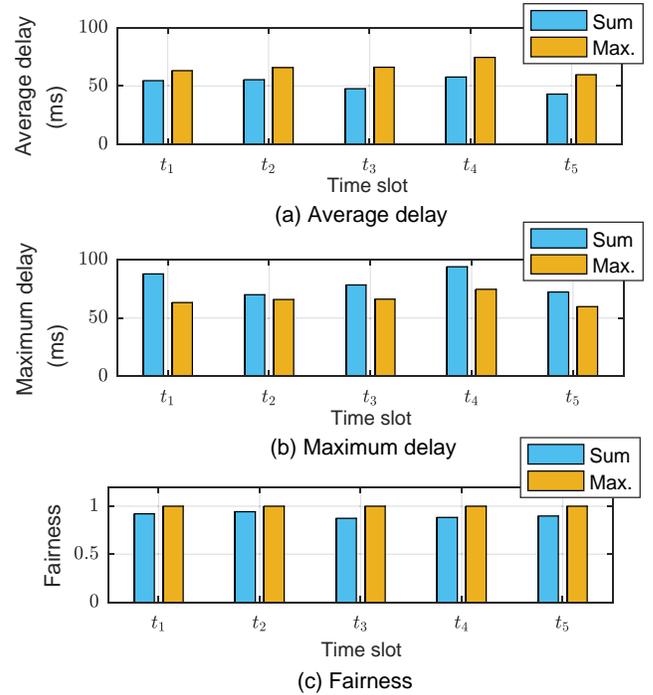}}
	\caption{Average delay, maximum delay, and fairness of the proposed scheme.}
	\label{fig14}
\end{figure}

Fig.~\ref{fig4} shows the convergence performance of Algorithm 1 in solving the sum-delay minimization problem (labeled as `Sum') and the maximum-delay minimization problem (labeled as `Max.') in our proposed scheme, where the delay performance of the `Sum' problem is measured by the average delay of all ICVs, and the `Max.' problem is measured by the maximum-delay value of all ICVs. As we can see, the performance of both delays converge to stable points within ten iterations, thus confirming the convergence and efficiency of the SCA method. 

	Fig.~\ref{fig14} (a), (b), and (c) respectively show the ICVs' average delay, maximum delay, and fairness index of the proposed scheme under the premise of solving the `Sum' problem and the `Max.' problem in five consecutive time slots. In this work, the fairness is measured by the widely used Jain's equation, defined by $\mathcal{J} =(\sum_{n=1}^{N} y_n)^2/(N\times\sum_{n=1}^{N}y_n^2)$. As we can see, the average delay of the `Sum' problem is better than that of the `Max.' problem, but its maximum delay value is worse. Under the `Max.' problem, the ICVs can obtain higher fairness values. This is because the goal of the `Sum' problem is to optimize the average delay performance of the ICVs, while the purpose of the `Max.' problem is to optimize the maximum delay of the ICVs.

\begin{figure}[!t]
	\centerline{\includegraphics[height=6.5cm]{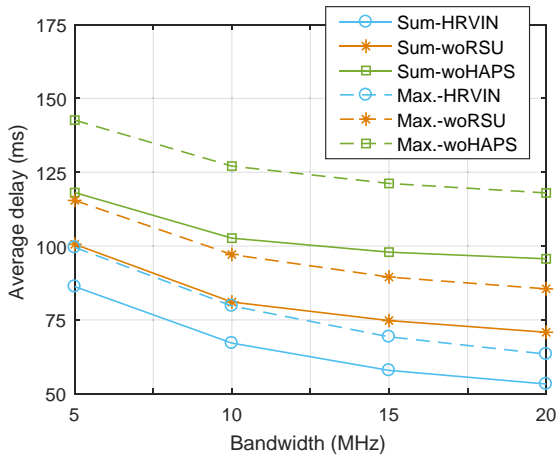}}
	\caption{Average delay vs. bandwidth.}
	\label{fig6}
\end{figure}

\begin{figure}[!t]
	\centerline{\includegraphics[height=6.5cm]{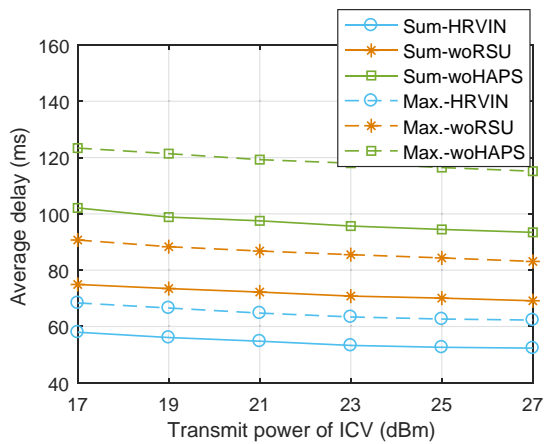}}
	\caption{Average delay vs. transmit power of ICV.}
	\label{fig7}
\end{figure}
\begin{figure}[!t]
	\centerline{\includegraphics[height=6.5cm]{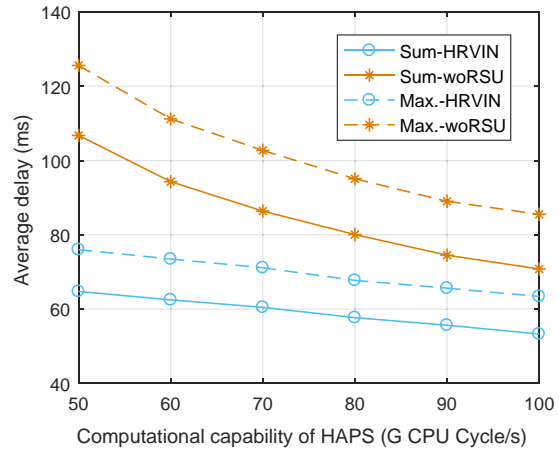}}
	\caption{Average delay vs. computational capability of HAPS.}
	\label{fig8}
\end{figure}
\begin{figure}[!t]
	\centerline{\includegraphics[height=6.5cm]{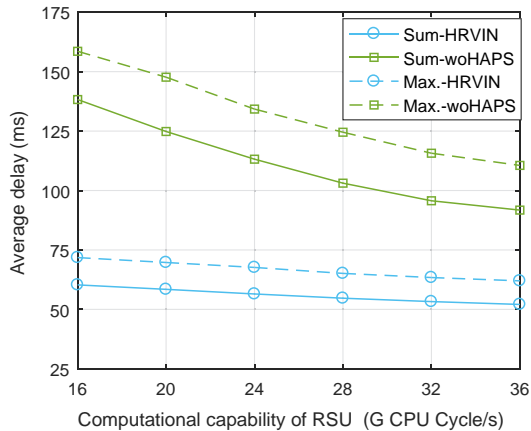}}
	\caption{Average delay vs. computational capability of RSU.}
	\label{fig9}
\end{figure}

Fig.~\ref{fig6} – Fig.~\ref{fig9} show the effects of bandwidth capacity, ICV's transmit power threshold, HAPS computational capability, and RSU computational capability settings on the average delay performance of different schemes when solving the `Sum' and `Max.' problems.  To distinguish these schemes, we use the label `HRVIN' to represent the proposed scheme within the distributed computing framework of the HAPS-RSU-vehicle integrated network, where the task of each ICV can be processed at its onboard device, RSU server and HAPS server in parallel. We use the label `woRSU' to represent the  two-layer computing scheme where there is no RSU computing, and we use the label `woHAPS' to represent the two-layer computing scheme where there is no HAPS computing, which can also be regarded as a traditional distributed computing scheme.  



Fig.~\ref{fig6} shows the impact of different bandwidth capacity settings on delay performance. As we can see, the `HRVIN' scheme with three layers of computing resources achieves better performance compared to the other two baseline schemes, which have only two layers of computing resources. This shows that the `HRVIN' scheme is more conducive to accelerating the task computing. Besides, by comparing the `woRSU' and `woHAPS' schemes, we can see that only the HAPS-assisted computing scheme `woRSU' can also effectively improve the delay performance greater than the 'woHAPS' scheme, and this is because the HAPS can be equipped with more computing resources than the RSU. Moreover, the increase of bandwidth capacity can reduce the delay of computing tasks. The reason for this is that the increase of bandwidth enables ICVs to  improve their ability to offload data to the RSUs and HAPS. Benefiting from this, more data can be offloaded to speed up the data processing, thereby improving the ICVs' perceived delays.

Fig.~\ref{fig7} shows the impact of different transmit power threshold settings on delay performance. As we can see, increasing the power threshold can reduce the delay. Similar to the change of bandwidth capacity, the increase in power threshold improves the ICV's ability to offload data.  This allows ICVs to offload more data for processing, thus improving the delay performance. 

Fig.~\ref{fig8} shows the impact of different  computational capability settings of the HAPS server on the delay performance. We can observe that with the increase of the computational capability, the delay can be gradually reduced. This suggests that the increase in the computational capability of the HAPS server can help ICVs to offload more data to the HAPS to obtain more powerful computing resources, thus accelerating the data processing.

Fig.~\ref{fig9} shows the impact of different  computational capability settings of the RSU server on the delay performance. On the whole, the increase in the server's computational capability can speed up the execution of tasks. 

\begin{figure}[!t]
	\centerline{\includegraphics[height=9cm]{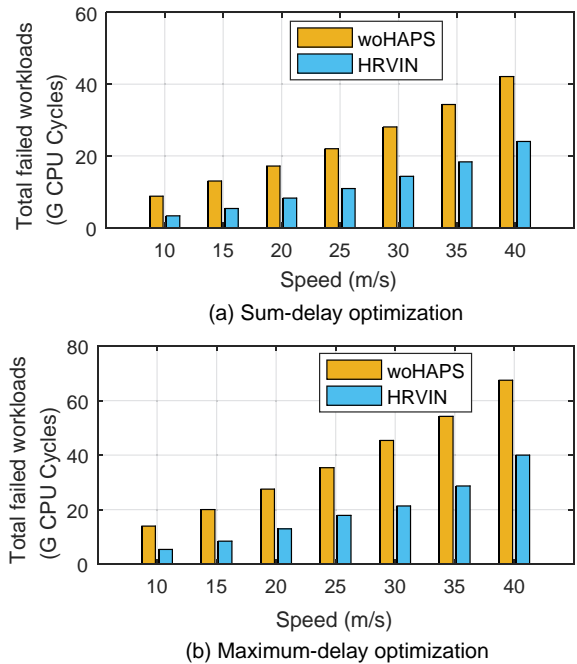}}
	\caption{The total failed workloads caused by a handoff vs. speed of ICV.}
	\label{fig10}
\end{figure}

\begin{figure}[!t]
	\centerline{\includegraphics[height=8.5cm]{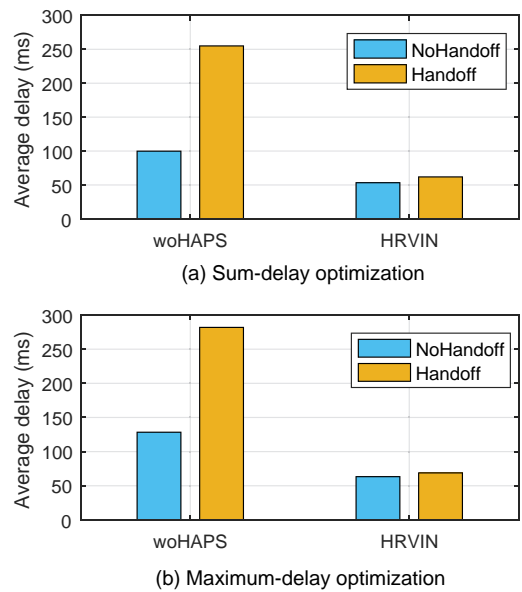}}
	\caption{The effect of handoffs on delay performance. }
	\label{fig11}
\end{figure}

As mentioned above, network handoffs between ICVs and RSUs will negatively affect computation offloading, and this problem will be more severe for high-speed mobile scenarios. To illustrate the impact of a network handoff and mobility on computing, Fig.~\ref{fig10} (a) and (b) show the total failed workloads of ICVs under different speed settings when removing network handoff constraint $T_n^R \le T_n^{handoff},\forall{n}\in\mathcal{N}$, for solving the sum-delay optimization problem, and the maximum-delay optimization problem, respectively. We count the averaged one-minute failed workloads, where the workload is defined as the product of the input data bits $\varepsilon$ (bit) and the computational density $\lambda$ (CPU Cycle/bit). As both figures show, the total failed workloads of ICVs increase as the speed increases on the whole. According to equation $T_n^{handoff}=\frac{D-|l_n~mod ~D|}{v}$, the handoff time is inversely proportional to the speed of the ICV. This means that with the increase of speed, the time to trigger the network handoff will be earlier, so the frequency of handoffs occurring during the driving will also increase, and the cumulative failed workloads will increase. Fig.~\ref{fig10} shows that considering the network handoff factor when designing a computing strategy can effectively avoid failures caused by handoff interruptions. Obviously, this is of great significance for efficient and successful data processing, especially for high-speed mobile scenarios. In addition, by comparing the `woHAPS' and `HRVIN' schemes, we can see that the failed workloads of the `woHAPS' scheme are higher than that of `HRVIN' scheme. The reason is that when we do not consider the handoff factor, most of the ICVs in the `woHAPS' scheme mainly rely on RSU computing, so the failed workloads caused by handoffs exceed the `HRVIN' scheme that can depend on both RSU computing and HAPS computing. Although considering network handoffs in the `woHAPS' scheme (i.e., adding the handoff constraint at RSUs) can avoid computational interruptions, it will force ICVs that encounter handoffs to rely only on their own local computing, thus resulting in poor delay performance. This fact will be verified in Fig.~\ref{fig11}. In comparison,  the `HRVIN' scheme that does not consider the network handoff factor reduces the computational burden on the RSU server to a certain extent. Therefore, when encountering a handoff, the interrupted workloads are fewer.
Further, if the handoff factor can be considered in the `HRVIN' scheme when encountering a handoff, a large amount of data can be offloaded to the HAPS, which can yield a lower delay compared with the `woHAPS' scheme that can only rely on local computing, and in summary, the negative impact of network handoffs on computing is avoided.  Additionally, HAPS computing can help to improve the efficiency of task computing significantly while avoiding the adverse effects of network handoffs. 

Fig.~\ref{fig11} shows the average delays of two cases, one where no handoff occurs and one where the handoff does occur, indicated by `NoHandoff' and `Handoff', respectively. In both sub-figures, we compare the delay performance of the `woHAPS' and `HRVIN' schemes to illustrate the importance of HAPS computing to the handoff case. As we can see, in each sub-figure, the edge computing scheme `woHAPS' has a delay increase of nearly 150 ms in the `Handoff' case compared to  the `NoHandoff' case, while the delay increase of the `HRVIN' scheme is only less than 10 ms. This is because, in order to avoid a computation interruption caused by a network handoff, the ICV in the `woHAPS' scheme can only offload a small amount of data to the edge, leaving most of the data to be computed locally, thus resulting in a significant increase in delay. However, in the `HRVIN' scheme, the ICV can send the data that the RSU cannot handle to the HAPS, so the delay increase is slight and acceptable. From the above comparison, we can see that the delay performance of traditional edge computing in the handoff case will suddenly deteriorate, which is obviously not conducive to the stability and safety of ICV driving. By contrast, a HAPS can resolve the negative impact of network handoff by completing task computing within an acceptable increase in delay. Therefore, introducing HAPS computing can help ICVs cope with the network handoffs.
\begin{figure}[!t]
	\vspace{7pt}
	\centerline{\includegraphics[width=10cm]{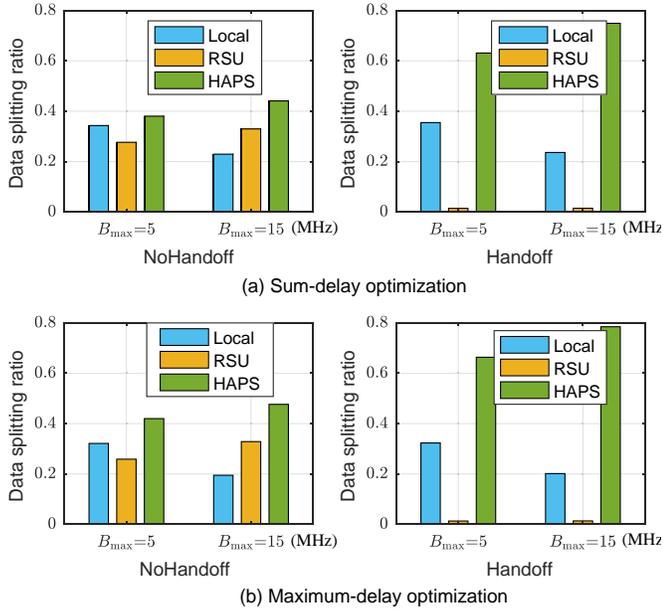}}
	\caption{The data splitting ratio for the NoHandoff and Handoff cases vs. bandwidth.}
	\label{fig12}
\end{figure} 
\begin{figure}[!t]
	\centerline{\includegraphics[width=10cm]{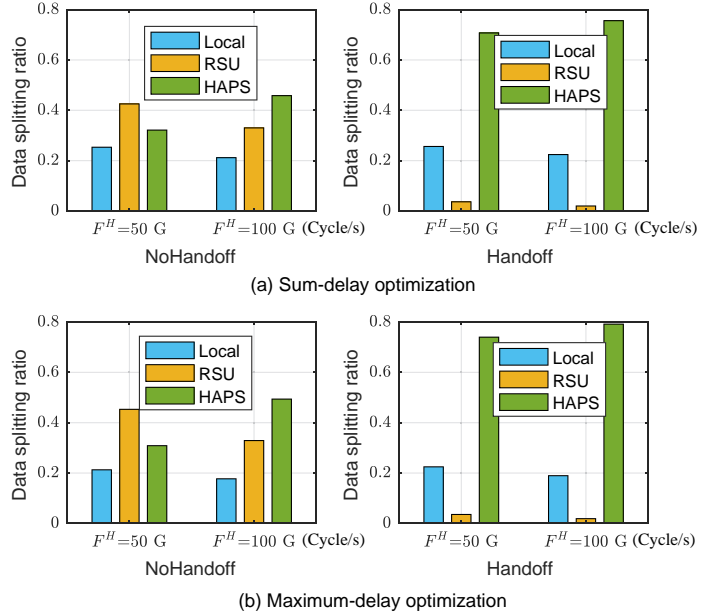}}
	\caption{The data splitting ratio for the NoHandoff and Handoff cases vs. computational capability of the HAPS server.}
	\label{fig13}
\end{figure}

Fig.~\ref{fig12} shows the averaged ratio of a task performed in three computing nodes: local, RSU, and HAPS with different bandwidth settings for the proposed scheme. Fig. ~\ref{fig12} (a) and (b) show the comparison of data splitting in the two situations of `NoHandoff' and `Handoff' with the premise of solving the sum-delay optimization problem. Fig. ~\ref{fig12} (c) and (d) reflect the above situations of solving the maximum-delay optimization problem. As we can observe in the `NoHandoff' case in sub-figures (a) and (b), when the bandwidth is 5 MHz, the data is mostly processed locally or on the HAPS server. This is because the small bandwidth limits the ability of ICVs to offload data. Consequently, the data tends to be directly processed by the ICV's onboard device, or offloaded to the HAPS that has more computing resources. As the bandwidth capacity increases to 15 MHz, the throughput of offloading data to the RSUs and HAPS significantly improves, effectively reducing the corresponding communication delay, so the splitting ratio of local computing decreases, and the splitting ratios of RSU computing and HAPS computing increase. In addition, when we compare the `NoHandoff' and `Handoff' cases, we can observe that the splitting ratio of RSU computing significantly drops from 25\%-35\% to 3\%. This is because the data offloaded to the RSUs needs to be successfully processed  before the network handoff is triggered. Hence, in the `Handoff' case, the data splitting ratio of the RSU is small. Meanwhile, the splitting ratio of the HAPS increases significantly from 38\%-46\% to 62\%-78\%, indicating that most of the RSU's workload has migrated to the HAPS to avoid computation interruptions. This tells us that HAPS computing plays an important role when the handoff occurs. 

Similarly, Fig.~\ref{fig13} shows the averaged ratio of a task performed in three computing nodes: local, RSU, and HAPS under different computational capability settings of the HAPS for the proposed scheme. As we can observe in the `NoHandoff' case in sub-figures (a) and (c), RSU computing plays a crucial role when the computational capability of the HAPS is set to 50 G CPU Cycle/s. This is because the computational capacity of the HAPS is smaller than the total capabilities of the two RSUs (each with 32 G CPU Cycle/s) in the system. When the computing capacity of the HAPS server increases to 100 G CPU Cycle/s, it is clear that HAPS computing plays a significant role. In addition, by comparing the `NoHandoff' and `Handoff' cases, we can ascertain that the HAPS takes on the majority of the workloads when the handoff occurs (i.e., about 70\%-80\%), while the RSU computing only handles a fraction of the workload. By looking at Fig.~\ref{fig12} and Fig.~\ref{fig13} together, we can draw two conclusions: First, in the `Handoff' case, most of the data has migrated from the RSU to the HAPS, so the HAPS can play a crucial role in this case, which confirms that HAPS computing in this distributed framework can effectively deal with network handoffs. Second, according to the data splitting ratios, the proposed distributed computing scheme can indeed achieve more flexible and adaptive task scheduling.
\section{Conclusion}
  In this work, we proposed a distributed parallel computing scheme with the assistance of HAPS to achieve a lower delay performance of vehicular computing tasks and at the same time provide a smooth and stable service experience for vehicles by avoiding the negative impact of handoff. As we saw, the scheme  flexibly divides data into three parts and  processes it in parallel on ICVs, RSUs, and a HAPS. By setting the task portion that is computed at the RSU to be completed before the network handoff, the scheme effectively eliminates the negative impact of the handoff on the computation. On this basis, this work formulated a total delay optimization problem and solved it using the SCA method. Then, we also discussed the formulation and solution of the maximum-delay optimization problem. Finally, extensive simulation results confirmed the effectiveness of our proposed scheme.  \par 
The energy consumption issue for HAPS network is critical because both hovering and computation require energy consumption. The role of HAPS in computing will be weakened when considering the energy issue of HAPS, which will make the delay perceived by ICVs longer.  In the future work, we will discuss this comprehensive and interesting topic. 
  
\bibliographystyle{ieeetr}
\bibliography{IEEEabrv,ICCworkshopJournal}

\begin{thebibliography}{10}

\bibitem{vcompu1}
J.~Zhou, D.~Tian, Y.~Wang, Z.~Sheng, X.~Duan, and V.~C. Leung,
  ``Reliability-optimal cooperative communication and computing in connected
  vehicle systems,'' {\em {IEEE} Trans. Mobile Comput.}, vol.~19,
  pp.~1216--1232, May 2020.

\bibitem{vehicle4}
K.~N. Qureshi, S.~Din, G.~Jeon, and F.~Piccialli, ``Internet of vehicles: Key
  technologies, network model, solutions and challenges with future aspects,''
  {\em {IEEE} Trans. Intell. Transp. Syst.}, vol.~22, pp.~1777--1786, Mar.
  2021.

\bibitem{mingzhechen1}
M.~Chen, Z.~Yang, W.~Saad, C.~Yin, H.~V. Poor, and S.~Cui, ``A joint learning
  and communications framework for federated learning over wireless networks,''
  {\em {IEEE} Trans. Wireless Commun.}, vol.~20, pp.~269--283, Jan. 2021.

\bibitem{vehicle2}
Z.~Ning, P.~Dong, X.~Wang, M.~S. Obaidat, X.~Hu, L.~Guo, Y.~Guo, J.~Huang,
  B.~Hu, and Y.~Li, ``When deep reinforcement learning meets {5G}-enabled
  vehicular networks: A distributed offloading framework for traffic big
  data,'' {\em {IEEE} Trans. Ind. Informat.}, vol.~16, pp.~1352--1361, Feb.
  2020.

\bibitem{jsac_chenmingzhe}
M.~Chen, D.~Gündüz, K.~Huang, W.~Saad, M.~Bennis, A.~V. Feljan, and H.~V.
  Poor, ``Distributed learning in wireless networks: Recent progress and future
  challenges,'' {\em {IEEE} J. Sel. Areas Commun.}, vol.~39, pp.~3579--3605,
  Dec. 2021.

\bibitem{editor3}
J.~Kang, X.~Li, J.~Nie, Y.~Liu, M.~Xu, Z.~Xiong, D.~Niyato, and Q.~Yan,
  ``Communication-efficient and cross-chain empowered federated learning for
  artificial intelligence of things,'' {\em {IEEE} Trans. Netw. Sci. Eng.},
  vol.~9, pp.~2966--2977, May 2022.

\bibitem{reviewer4}
Y.~Ye, R.~Q. Hu, G.~Lu, and L.~Shi, ``Enhance latency-constrained computation
  in {MEC} networks using uplink {NOMA},'' {\em {IEEE} Trans. Commun.},
  vol.~68, pp.~2409--2425, Apr. 2020.

\bibitem{mingzhechen2}
M.~Chen, N.~{Shlezinger}, , H.~V. Poor, and S.~Cui, ``Communication efficient
  federated learning,'' {\em Proceedings of the National Academy of Sciences of
  the United States of America}, vol.~118, Apr. 2021.

\bibitem{Car}
{NTT DATA White Paper}, ``When the car takes over: a glimpse into the future of
  autonomous driving.'' Accessed: Jun. 10, 2022. [Online]. Available:
  \url{https://us.nttdata.com/en/-/media/assets/white-paper/mfg-autonomous-cars-white-paper.pdf}.

\bibitem{vehicle3}
F.~Lyu, H.~Zhu, N.~Cheng, H.~Zhou, W.~Xu, M.~Li, and X.~Shen, ``Characterizing
  urban vehicle-to-vehicle communications for reliable safety applications,''
  {\em {IEEE} Trans. Intell. Transp. Syst.}, vol.~21, pp.~2586--2602, Jun.
  2019.

\bibitem{V2V}
G.~Qiao, S.~Leng, K.~Zhang, and Y.~He, ``Collaborative task offloading in
  vehicular edge multi-access networks,'' {\em {IEEE} Commun. Mag.}, vol.~56,
  pp.~48--54, Aug. 2018.

\bibitem{UAV2V}
W.~Shi, H.~Zhou, J.~Li, W.~Xu, N.~Zhang, and X.~Shen, ``Drone assisted
  vehicular networks: Architecture, challenges and opportunities,'' {\em {IEEE}
  Netw.}, vol.~32, pp.~130--137, Jun. 2018.

\bibitem{Sate_IOV}
M.~LiWang, S.~Dai, Z.~Gao, X.~Du, M.~Guizani, and H.~Dai, ``A computation
  offloading incentive mechanism with delay and cost constraints under {5G}
  satellite-ground {IoV} architecture,'' {\em {IEEE} Wireless Commun.},
  vol.~26, pp.~124--132, Aug. 2019.

\bibitem{editor1}
H.~Du, D.~Niyato, Y.-A. Xie, Y.~Cheng, J.~Kang, and D.~I. Kim, ``Performance
  analysis and optimization for jammer-aided multiantenna {UAV} covert
  communication,'' {\em {IEEE} J. Sel. Areas Commun.}, vol.~40, pp.~2962--2979,
  Oct. 2022.

\bibitem{HAPcomputing4}
Z.~Jia, M.~Sheng, J.~Li, and Z.~Han, ``Toward data collection and transmission
  in {6G} space–air–ground integrated networks: Cooperative {HAP} and {LEO}
  satellite schemes,'' {\em {IEEE} Internet Things J.}, vol.~9,
  pp.~10516--10528, Jul. 2022.

\bibitem{reviewer2}
Z.~Jia, Q.~Wu, C.~Dong, C.~Yuen, and Z.~Han, ``Hierarchical aerial computing
  for internet of things via cooperation of {HAPs} and {UAVs},'' {\em {IEEE}
  Internet Things J.}, 2022.
\newblock Early Access.

\bibitem{ITU}
ITU, ``Radio regulations articles.''
  \url{http://www.itu.int/pub/R-REG-RR-2016}, 2016.

\bibitem{reviewer1}
Z.~Jia, M.~Sheng, J.~Li, D.~Zhou, and Z.~Han, ``Joint {HAP} access and {LEO}
  satellite backhaul in {6G}: Matching game-based approaches,'' {\em {IEEE} J.
  Sel. Areas Commun.}, vol.~39, pp.~1147--1159, Apr. 2021.

\bibitem{WaelITS}
W.~{Jaafar} and H.~{Yanikomeroglu}, ``{HAPS-ITS}: Enabling future {ITS}
  services in trans-continental highways,'' {\em {IEEE} Commun. Mag.}, vol.~60,
  pp.~80--86, Oct. 2022.

\bibitem{wind}
``The promise and challenges of airborne wind energy.'' Accessed: May 18, 2021.
  [Online]. Available:
  \url{https://physicsworld.com/a/the-promise-and-challenges-of-airborne-wind-energy/}.

\bibitem{battery}
``{HAPSMobile} and {APB} reach basic agreement to develop storage batteries for
  {HAPS} using all polymer battery.'' Accessed: May 18, 2021. [Online].
  Available: \url{https://www.hapsmobile.com/en/news/press/2020/20201224_01/}.

\bibitem{Gunes1}
G.~K. {Kurt}, M.~G. {Khoshkholgh}, S.~{Alfattani}, A.~{Ibrahim}, T.~S.~J.
  {Darwish}, M.~S. {Alam}, H.~{Yanikomeroglu}, and A.~{Yongacoglu}, ``A vision
  and framework for the high altitude platform station ({HAPS}) networks of the
  future,'' {\em {IEEE} Commun. Surveys Tuts.}, vol.~23, pp.~729--779,
  Secondquarter 2021.

\bibitem{HAPcomputing}
Z.~Jia, Q.~Wu, C.~Dong, C.~Yuen, and Z.~Han, ``Hierarchical aerial computing
  for internet of things via cooperation of {HAPs} and {UAVs},'' {\em {IEEE}
  Internet Things J.}, 2022.
\newblock (Early Access).

\bibitem{Gunes2}
G.~K. {Kurt} and H.~{Yanikomeroglu}, ``Communication, computing, caching, and
  sensing for next-generation aerial delivery networks: Using a high-altitude
  platform station as an enabling technology,'' {\em {IEEE} Veh. Technol.
  Mag.}, vol.~16, pp.~108--117, Sep. 2021.

\bibitem{QiqiITS}
Q.~{Ren}, O.~{Abbasi}, G.~K. {Kurt}, H.~{Yanikomeroglu}, and J.~{Chen},
  ``Caching and computation offloading in high altitude platform station
  ({HAPS}) assisted intelligent transportation systems,'' {\em {IEEE} Trans.
  Wireless Commun.}, vol.~21, pp.~9010--9024, Nov 2022.

\bibitem{ShuaiyuSAGIN}
S.~Yu, X.~Gong, Q.~Shi, X.~Wang, and X.~Chen, ``{EC-SAGINs}:
  Edge-computing-enhanced space-air-ground-integrated networks for internet of
  vehicles,'' {\em {IEEE} Internet Things J.}, vol.~9, pp.~5742--5754, Apr.
  2022.

\bibitem{hand40}
V.~B. Souza, M.~H. Pereira, L.~H.~S. Lelis, and X.~Masip-Bruin, ``Enhancing
  resource availability in vehicular fog computing through smart inter-domain
  handover,'' in {\em GLOBECOM 2020 - 2020 IEEE Global Communications
  Conference}, pp.~1--6, Dec. 2020.

\bibitem{hand41}
T.~M. Ho and K.-K. Nguyen, ``Joint server selection, cooperative offloading and
  handover in multi-access edge computing wireless network: A deep
  reinforcement learning approach,'' {\em {IEEE} Trans. Mobile Comput.},
  vol.~21, pp.~2421--2435, Jul. 2022.

\bibitem{hand43}
Q.~Yuan, J.~Li, H.~Zhou, T.~Lin, G.~Luo, and X.~Shen, ``A joint service
  migration and mobility optimization approach for vehicular edge computing,''
  {\em {IEEE} Trans. Veh. Technol.}, vol.~69, pp.~9041--9052, Aug. 2020.

\bibitem{hand44}
W.~Zhan, C.~Luo, J.~Wang, C.~Wang, G.~Min, H.~Duan, and Q.~Zhu,
  ``Deep-reinforcement-learning-based offloading scheduling for vehicular edge
  computing,'' {\em {IEEE} Internet Things J.}, vol.~7, pp.~5449--5465, Jun.
  2020.

\bibitem{hand45}
H.~Zhang, R.~Wang, W.~Sun, and H.~Zhao, ``Mobility management for
  blockchain-based ultra-dense edge computing: A deep reinforcement learning
  approach,'' {\em {IEEE} Trans. Wireless Commun.}, vol.~20, pp.~7346--7359,
  Nov. 2021.

\bibitem{hand46}
T.~Ojima and T.~Fujii, ``Resource management for mobile edge computing using
  user mobility prediction,'' in {\em 2018 International Conference on
  Information Networking (ICOIN)}, pp.~718--720, Jan. 2018.

\bibitem{JsacHand}
C.~W. Lee, L.~M. Chen, M.~C. Chen, and Y.~S. Sun, ``A framework of handoffs in
  wireless overlay networks based on mobile {IPv6},'' {\em {IEEE} J. Sel. Areas
  Commun.}, vol.~23, pp.~2118--2128, Nov. 2005.

\bibitem{2007An}
M.~Liu, Z.~C. Li, and X.~B. Guo, ``An efficient handoff decision algorithm for
  vertical handoff between {WWAN} and {WLAN},'' {\em Journal of Computer
  Science and Technology}, vol.~22, pp.~114--120, Jan. 2007.

\bibitem{JSAC_hand2}
K.~Shafiee, A.~Attar, and V.~C.~M. Leung, ``Optimal distributed vertical
  handoff strategies in vehicular heterogeneous networks,'' {\em {IEEE} J. Sel.
  Areas Commun.}, vol.~29, pp.~534--544, Mar. 2011.

\bibitem{TWC1}
A.~{Ibrahim} and A.~S. {Alfa}, ``Using {Lagrangian} relaxation for radio
  resource allocation in high altitude platforms,'' {\em {IEEE} Trans. Wireless
  Commun.}, vol.~14, pp.~5823--5835, Jun. 2015.

\bibitem{editor2}
J.~Kang, H.~Du, Z.~Li, Z.~Xiong, S.~Ma, D.~Niyato, and Y.~Li, ``Personalized
  saliency in task-oriented semantic communications: Image transmission and
  performance analysis,'' {\em {IEEE} J. Sel. Areas Commun.}, vol.~41,
  pp.~186--201, Nov. 2023.

\bibitem{TWC2}
A.~{Alsharoa} and M.~S. {Alouini}, ``Improvement of the global connectivity
  using integrated satellite-airborne-terrestrial networks with resource
  optimization,'' {\em {IEEE} Trans. Wireless Commun.}, vol.~19,
  pp.~5088--5100, Apr. 2020.

\bibitem{HAPsurvey2005}
S.~{Karapantazis} and F.~{Pavlidou}, ``Broadband communications via
  high-altitude platforms: A survey,'' {\em IEEE Communications Surveys
  Tutorials}, vol.~7, pp.~2--31, Firstquarter 2005.

\bibitem{cvx}
S.~Boyd and L.~Vandenberghe, {\em Convex Optimization}.
\newblock Cambridge University Press.

\bibitem{b4}
J.~Kaleva, A.~Tölli, and M.~Juntti, ``Decentralized sum rate maximization with
  {QoS} constraints for interfering broadcast channel via successive convex
  approximation,'' {\em {IEEE} Trans. Signal Process.}, vol.~64,
  pp.~2788--2802, Jun. 2016.

\bibitem{b5}
A.~Liu, V.~K.~N. Lau, and M.~Zhao, ``Online successive convex approximation for
  two-stage stochastic nonconvex optimization,'' {\em {IEEE} Trans. Signal
  Process.}, vol.~66, pp.~5941--5955, Nov. 2018.

\bibitem{Renqiqi2}
Q.~{Ren}, J.~{Chen}, O.~{Abbasi}, G.~K. {Kurt}, H.~{Yanikomeroglu}, and F.~R.
  {Yu}, ``An application-driven non-orthogonal multiple access enabled
  computation offloading scheme,'' {\em {IEEE} Internet Things J.}, vol.~8,
  pp.~1453--1466, Feb. 2021.

\bibitem{b50}
Q.~{Wu}, Y.~{Zeng}, and R.~{Zhang}, ``Joint trajectory and communication design
  for multi-{UAV} enabled wireless networks,'' {\em {IEEE} Trans. Wireless
  Commun.}, vol.~17, pp.~2109--2121, Mar. 2018.

\bibitem{OmidTWC}
O.~{Abbasi}, H.~{Yanikomeroglu}, A.~{Ebrahimi}, and N.~M. {Yamchi},
  ``Trajectory design and power allocation for drone-assisted {NR-V2X} network
  with dynamic {NOMA/OMA},'' {\em {IEEE} Trans. Wireless Commun.}, vol.~19,
  pp.~7153--7168, Nov. 2020.

\bibitem{17dbi}
F.~B. {Mismar} and B.~L. {Evans}, ``Partially blind handovers for {mmWave} new
  radio aided by sub-6 {GHz} {LTE} signaling,'' in {\em 2018 IEEE International
  Conference on Communications Workshops (ICC Workshops)}, pp.~1--5, May 2018.

\end{thebibliography}

\end{document}